\documentclass[aps,prl,twocolumn,superscriptaddress,
amsfonts,amssymb,amsmath,
twoside,
letterpaper,
bibnotes,
]{revtex4}

\pdfoutput=1

\usepackage{bm,color,mathrsfs,graphicx}

\begin{document}

\title{
Pseudo-proper ferroelectricity in thin films
}

\author{A. Cano}
\email{cano@esrf.fr}
\affiliation{
European Synchrotron Radiation Facility, 6 rue Jules Horowitz, BP 220, 38043 Grenoble, France}

\author{A.P. Levanyuk}
\affiliation{
European Synchrotron Radiation Facility, 6 rue Jules Horowitz, BP 220, 38043 Grenoble, France}
\affiliation{
\mbox{Departamento de F\' isica de la Materia Condensada, C-III, Universidad Aut\' onoma de Madrid, E-28049 Madrid, Spain}}

\date{\today}

\begin{abstract} 
We study ferroelectricity in thin films of pseudo-proper ferroelectrics such as the so-called spiral multiferroics. We find that this type of ferroelectricity stands better against depolarizing fields than conventional one. Its single-domain state can be easily preserved by metallic electrodes even in ultrathin films. In fact, single-domain ferroelectricty can be generated as a metastable state in the absence of electrodes. We also find a new regime of small thickness where unscreened films develop unusual multi-domain states with properties determined by non-electrostatic boundary conditions. 
\end{abstract}


\maketitle

\paragraph{Introduction.--} The discovery of a new type of ferroelectricity caused by a cycloidal ordering of magnetic moments \cite{Kimura03} is one of the key events that has triggered the current interest in multiferroics. This type of multiferroicity is realized in the rare-earth manganites $R$MnO$_3$ ($R=$ Gd, Tb, Dy), and has been successfully explained on the basis of the inverse Dzyaloshinskii-Moriya interaction \cite{Katsura05}. From the phenomenological point of view, these systems can be considered as new examples of pseudo-proper ferroelectrics \cite{StrukovLevanyuk} in which the electric polarization is (bi-)linearly coupled with the primary order parameter of the transition \cite{Cano08}.

Much of the interest in ferroelectrics concerns their thin-film properties
since they are at the root of a large number of applications (in memory devices, field-effect transistors, etc.). For conventional proper ferroelectrics this has been addressed theoretically by means of both first-principles calculations \cite{Junquera08} and Landau-like approaches \cite{Bratkovsky09}. One of the main outcomes is that single-domain (uniform) ferroelectricity is very difficult to retain below certain sizes. Instead, there appears multi-domain structures which are largely the result of the depolarizing fields that invebitably persist in real devices \cite{fnote1}. 
As regards pseudo-proper ferroelectrics, the thin film properties of these systems have not yet been addressed in the same detail. 

In this paper we show that pseudo-proper ferroelectricity proves more robust against depolarizing field effects. 
We illustrate this robustness in different experimental situations. 
When the film is sanwiched between metallic short-circuited electrodes, for example, the critical screening length necesary to keep single-domain ferroelectricity is much larger than in conventional ferroelectrics. In TbMnO$_3$, for example, it is expected $\sim150$\AA, which largely exceeds the screening length of many metals. 
The physics behind this result has nothing to do with any eventual smallness in the induced polarization. This can be noted in a convential ferroelectric, where no increase of the critical screening length is obtained by the mere reduction of its spontaneous polarization. 
The reason is that the aforementioned linear coupling sets a new length scale $\lambda$ in the problem, and this quantity overwhelms the typical length scale for the gradients of polarization in determining the above critical screening. 
On the other hand, in the absence of electrodes single-domain ferroelectricity can be generated as a metastable state in striking contrast to conventional ferroelectrics. 
This is possible because the pseudo-proper mechanism does not require the vanishing of the polarization stiffness. Therefore, despite the depolarizing field increases this stifness, the virtual single-domain ferroelectric instability remains sufficiently close in temperature. 
In addition, the new length scale $\lambda$ opens a small-thickness regime in which non-electrostatic boundary conditions become important for multi-domain ferroelectricity. This is evidenced in the period of the states that cause the instability of the paraelectric phase in a film without electrodes. For film thicknesses $l \gg \lambda$ the period decrease by reducing $l$ as in conventional ferroelectrics. When $l$ reaches $\lambda$, however, it becomes comparable to the film thickness, and therefore the average out of the corresponding depolarizing field ceases to be efficient. A further reduction of $l$ then produces an increase of such a period whose precise form depends on the non-electrostatic boundary conditions. 
This latter behavior, unlike in the magnetic case \cite{Garel82}, is rather unusual for a ferroelectric and, to the best of our knowledge, has been unnoticed so far. 

\paragraph{Equations of state.--} The instability towards ferroelectricity in the pseudo-proper case can be analyzed from the (linearized) equations of state:
\begin{subequations}\begin{align}
A P - f \eta &= - \partial_z V, \label{EqP}\\
\big(a - c _\parallel\partial_y^2  - c_\perp \partial_z^2 \big)\eta -fP&=0,\label{Eqeta}\\
(\varepsilon_y\partial_y^2 + \partial_z^2)V - 4\pi \partial_z P&=0.
\label{Max}
\end{align}\label{linearized}\end{subequations}
Here $P$ is the electric polarization (assumed to be perpendicular to the film), $V$ is the electrostatic potential and $\eta$ is the primary order parameter. The coefficient $a$ is therefore the control parameter, $a = a'(T-T_0)$, whereas $A$ represents the bare polarization stiffness, $f$ is the coupling constant that allows for ferroelectricity, $c_\parallel$ and $c_\perp$ account for the extra stiffness of non-uniform distributions of $\eta$, and $\varepsilon_y$ is the in-plane dielectric constant. 
Eq. \eqref{Max} results from Maxwell's equations and Eqs. \eqref{EqP} and \eqref{Eqeta} are derived in the appendix for the case of spiral multiferroics [where $\eta$ describes the transversal component of the magnetic cycloid: $\mathbf M =(0, \xi \cos Q y, \eta \sin Q y)$]
\cite{fnote2}.

The fact that $A$ does not vanish allows us to neglect derivatives of $P$ in Eq. \eqref{EqP}. Moreover, the essential physics due to depolarizing field effects can be revealed putting $c_\perp =0$. In this hypotetical case the order-parameter turns out to be insensitive to its boundary conditions (or, more physically, it can adjust to them within a zero-distance interval). This simplifies notably the algebra, and proves to describe correctly the behavior of the system for the case of the so-called natural boundary conditions $\partial_z \eta =0$. So, keeping in mind these restrictions, we take advantage of this model case to investigate the point at which the paraelectric phase losses its stability. That is, the point at which, by decreasing $a$, there appears the first nontrivial solution of the system of equations \eqref{linearized} that, in addition, satisfies the remaining (electrostatic) boundary conditions.

\paragraph{Perfect screening.--} First of all, it is convenient to revise the case in which the depolarizing field is completely screened. 
In that case, the stability is lost with respect to the single-domain state in which both $P$ and $\eta$ are constant. This happens at the critical value of the control parameter $a_c = f^2/A$, which corresponds to the critical temperature $T_c = T _0 + \Theta $ where $\Theta = f^2 / (a' A)$. The inverse susceptibility in the paraelectric phase then can be written as 
$\chi_e^{-1}  = A- {f^2  \over a } = A{(T-T_{c})/\Theta\over 1+ (T-T_{c})/\Theta}$.
This implies a Curie-Weiss behavior in the vicinity of the phase transition, as obtained from the first term in the expansion in powers of $T-T_{c}$:
$\chi_e^{-1}\approx A {(T-T_{c})/\Theta }$. Experimentally this behavior is observed only in a very narrow region around $T_{c}$ (see e.g. \cite{Kimura03}). This can be understood as a result of a small $\Theta$ that reveals that the coupling between $P$ and $\eta$ is effectively weak. In TbMnO$_3$, for example, it can be estimated $\Theta \lesssim 1\,$K ($T_c \sim 27\,$K in this case).

\paragraph{Partial screening.--} If the screening is not complete, there is a competition between the above uniform solution of the equations of state and multi-domain structures that vary periodically within the film plane. Close to the transition point these complicated structures reduce in practice to a single harmonic, i.e., $\sim e^{ik_y y}$. Thus, for a given $k_y$, the functions describing the corresponding structure can be sought in the form $\eta =\eta_0 e^{ik_y y}\cos k_z z$, $P = P_0(\eta_0) e^{ik_y y}\cos k_z z$ and $V = V_0(\eta_0) e^{ik_y y}\sin k_z z$ in view of the form of the equations \eqref{linearized} and the symmetry of the problem. Substituting in \eqref{linearized} one can see that the existence of such solutions implies the relation
\begin{align}
a = {f^2\over A + 4\pi {k_z^2 \over \varepsilon_y k_y^2 + k_z ^2}} - c_\parallel k_y^2 
\label{a_twowaves}\end{align}
between the parameters $k_y$, $k_z$ and $a$ (recall that $c_\perp =0$ for the while).

The largest values of the parameter $a$ are obtained for solutions with $k_z \ll k_y$. Physically this is because these structures produce the minimal electric field in the ferroelectric, and therefore have the lowest stiffness. Consequently these structures are the most natural suspects of being responsible for the instability of the paraelectric phase as they are in conventional ferroelectrics. 
To move on we have to check whether this is actually compatible with the corresponding boundary conditions.

\paragraph{Real electrodes.--} Consider first the case of a film sandwiched between short-circuited electrodes. We model the imperfect screening in the metal with insultating dead layers of thickness $d$ \cite{Bratkovsky09} and assume that the film thickness is $l\gg d$ in the following. Then, to satisfy the electrostatic boundary conditions, the above solutions have to be such that
\begin{align}
k_z  \tan {k_z l \over 2} = \varepsilon_y k_y \left( {k_y d \over 2} \right)
\label{}\end{align}
if $k_y d \ll 1$ as we expect. The maximun $a$ will be obtained for $k_z l \ll 1$, which is clearly compatible with $k_z \ll k_y$ if $d \ll l$. We then have \mbox{$k_y^2 = \left( 1 + {1\over 3}\left( {k_z l \over 2} \right)^2 +\dots\right){l\over \varepsilon_y d}k_z^2$}. 
Substituting in \eqref{a_twowaves} we can see that, below the critical dead-layer thickness
\begin{align}
d_c = \sqrt{3 A \over \pi \varepsilon_y }\lambda,
\label{}\end{align}
where $\lambda = (Ac)^{1/2}/|f|$, the single-domain state appears before these multi-domain structures for the value
\mbox{$a_c = {f^2 /(A + 4\pi {d\over l})}$} of the control parameter. 

As we see, $d_{c}$ is inversely proportional to the strength of the linear coupling between $P$ and $\eta$ which, ``by definition,'' has to be relatively small in pseudo-proper ferroelectrics. In TbMnO$_3$, for example, the estimates of the spin-phonon coupling given in \cite{Katsura07} indicate that $\lambda \sim 150$\AA. On the other hand $A, \varepsilon_y \sim 20-30$ in accordance with \cite{Kimura03}. Thus, in contrast to the conventional case \cite{Bratkovsky09}, the screening of practically any electrode will prevent the splitting of a pseudo-proper ferroelectric into different domains at the transition point.

\paragraph{No electrodes.--} The hampering of multi-domain states in pseudo-proper ferroelectrics is also manifested for dead layers thicker than $d_c$. Consider the extreme case in which there are no electrodes, i.e., there is no screening of the depolarizing field. In this case, the electrostatic boundary conditions are such that 
\begin{align}
k_z  \tan {k_z l \over 2} = \varepsilon_y k_y. 
\label{ebc_noelectrodes}\end{align}
for the multi-domain structures considered before. $k_z$ then has to be $\sim \pi/l $ to satisfy the condition $k_z \ll k_y$. Accordingly \mbox{$k_y \approx {2k_z \over \varepsilon_y(\pi - k_zl)}$} and, substituting into Eq. \eqref{a_twowaves}, we can see that these states do not appear before the single-domain solution if the thickness of the film is smaller than $\lambda$. One may then naively conclude that, below this thickness, the loss of stability takes place without the formation of domains since the multi-domain states that normally appear in conventional ferroelectrics get suppressed \cite{Cano09}. However this is not the end of the story, since less conventional structures may come into play. We have so far ruled out structures with $k_z \gtrsim k_y$ because they seem unfavorable from the point of view of the depolarizing field. This option, however, has to be reconsidered for $l<\lambda$. It corresponds to solutions with $k_zl\ll 1$, for which \mbox{$k_y = \left({k_z l \over 2\varepsilon_y}\right)\left( 1 + {1\over 3}\left( {k_z l \over 2} \right)^2 +\dots\right)k_z$} in accordance with \eqref{ebc_noelectrodes}. One can see that, in fact, these solutions appear before the single-domain state for $l<\lambda$, thus causing the instability of the paraelectric phase. 

\begin{figure}[t]
\includegraphics[width=.45\textwidth]{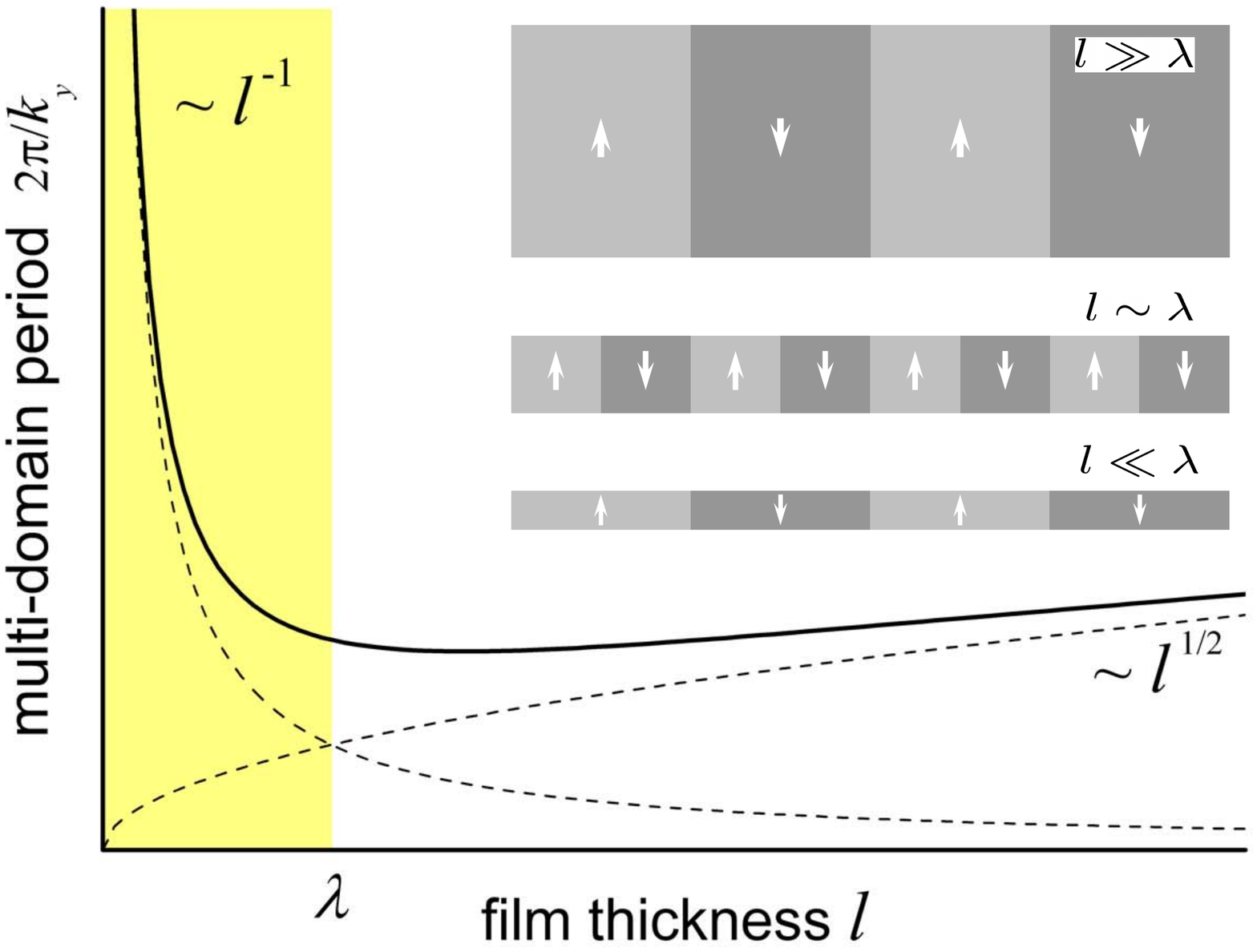}
\caption{Period of the multi-domain structures for an unscreened film of a pseudo-proper ferroelectric a function of its thickness. $\lambda$ is the characteristic length associated with the linear coupling between the polarization and the primary order parameter. Inset: schematic illustration of the corresponding behavior. 
}
\label{period}
\end{figure}

We note that the period of the conventional structures that appear for $l \gg \lambda $ varies $\sim (\lambda l)^{1/2}$ with the film thickness (see Fig. \ref{period}). This period becomes larger than the film thickness if $l<\lambda$, which explains the further tendency of the system to get rid of these structures: they are no longer effective in reducing locally the depolarizing field. Once this efficiency is lost, the factors that determine the subsequent behavior for $l \ll \lambda $ are mainly intrinsic (i.e., the linear coupling, gradient stiffness and non-electrostatic bondary conditions). Thus the period of the structures that appear in this regime varies unusually $\sim \lambda^2/ l$ (see Fig. \ref{period}). 
It is worth noting that conventional ferroelectrics only develop the former structures whose period increases with the film thickness, as will be the case of pseudo-proper ferroelectrics with extremely small $\lambda$'s unattainable experimentally (i.e., with strong couplings). 

This scenario, in which natural boundary conditions for $\eta$ are implicit, holds for $c_\perp \not =0$. The only changes are the following. The relation obtained instead of Eq. \eqref{a_twowaves} has two $k_z$ roots for a given $k_y$. Accordingly the general solution of the linearized equations of state is the linear combination of the corresponding functions. For $l > \lambda$ the second $k_z$ is associated with a surface contribution that gives exponentially small corrections to the structures described before. The paraelectric instability is therefore practically insensitive to the non-electrostatic boundary conditions as in conventional ferroelectrics. For $l < \lambda$, however, the second $k_z$ becomes relevant to describe the $z$-dependence of the structure that appears at the transition point. But this changes neither the period of the structures obtained above nor the fact that these solutions appear before the single-domain state. These latter results, however, are sensitive to the precise form of the non-electrostatic boundary conditions. 

We also note that pseudo-proper ferroelectrics have the follwing remarkable property. 
We have seen that, without electrodes, the instability of the paraelectic phase implies the appearance of multi-domain ferroelectricity. But the single-domain state is relatively close in energy in spite of its depolarizing field ($A + 4 \pi \simeq A$). Thus, by lowering the temperature, it is possible to have the situation in which the free energy develops at least a local minimum about the single-domain solution. That is, a situation in which single-domain ferroelectricity is metastable. This can be revealed by considering 
the tentative state of local equilibrium \mbox{$\eta_0^2 = -{1\over b}\big(a - {f^2\over A+ 4 \pi}\big)$}, where $b$ is the coefficient of the nonlinear term $b\eta^3$ omitted so far in \eqref{Eqeta}, and 
computing the stiffness associated with the perturbations $\eta_0 \to \eta_0 + \eta'$ that could drive the system out of that state. It is clear that, among all possible perturbations, those associated with the multi-domain structures considered before are the best candidates to do this job. But if the control parameter is $a \leq {f^2\over A + 4\pi}\left(1-{2\pi\over A}\right)$ the resulting stiffnesses are positive. The single-domain state is then robust against its splitting into different domains, even though the corresponding depolarizing field is not screened (which is unthikable in a conventional ferroelectric).

\paragraph{Conclusions.--} 
We have studied the specific features of the instability towards ferroelectricity in thin films of pseudo-proper ferroelectrics. Single-domain (uniform) ferroelectricity can be kept in films sandwiched between short-circuited electrodes without requiring any exceptional screening in the metal. In fact, this state can be generated as a metastable state even without electrodes. In addition, the properties of multi-domain states reveal new fundamental physics related to non-electrostatic boundary conditions.

We acknowledge P. Bruno, M. Civelli, E. Kats, and I. Paul for very fruitful discussions.

\onecolumngrid

\

\newpage 


\subsection{Appendix: Equations of state for spiral multiferroics}

In spiral multiferroics such as rare-earth manganites $R$MnO$_3$ ($R=$ Gd, Tb, Dy), ferroelectricity is associated with the magnetic phase transition from a longitudinal spin-density-wave to a cycloidal distribution of magnetization. In the following we parametrize the magnetization as $\mathbf M =(0, \xi \cos Q y, \eta \sin Q y)$ for the sake of simplicity. Therefore $\eta$ is the ``primary'' order parameter of the transition of our interest, while the concomitant polarization $\mathbf P$ is ``secondary'' in the sense that, whithout the coupling to $\eta$, it would remain zero. The relevant coupling is due to the inhomogeneous magnetoelectric effect \cite{Baryakhtar83}, which can been taken in its simplest form \cite{Mostovoy06}:
\begin{align}
F_{ME} &= f_0{\mathbf P}\cdot [({\mathbf M}\cdot \nabla ){\mathbf M} - {\mathbf M}(\nabla \cdot {\mathbf M}) ],
\label{}\end{align}
since we are interested in distributions of polarization that vary only at relatively large scales. 

To our purposes, it suffices to consider space variations along the $y$ and $z$ directions and the $z$-component of the polarization only 
\footnote{The nonuniform states that we consider also have $y$-component of the polarization that, however, is much smaller than the $z$ one that we consider explicitly.
}. 
Thus, neglecting for a while the depolarizing field,  the free energy of the paramagnetic phase can be taken as
\begin{align}
F &= {A \over 2}P^2 + {C\over 2}\left[(\partial_y P)^2 + (\partial_z P)^2  \right] 
+ f_0 P  \left[M_z (\partial_y M_y ) - M_y (\partial_y M_z) \right]
+{a_y \over 2}M_y^2 + {a_z \over 2}M_z^2 + {b\over 4}|\mathbf M|^4
\nonumber \\&\quad 
- {c_\parallel \over 2}|(\partial_y \mathbf M)|^2  + {g \over 4}|(\partial_y^2 \mathbf M)|^2 
+ {c_\perp }|(\partial_z\mathbf M)|^2.
\label{F_PM_inh}\end{align}

The appearance of the longitudinal magnetization wave $\mathbf M = (0, \xi \cos Qy,0)$ can be described by assuming that $c_\parallel > 0$. Thus, the wavevector of this modulation is $Q=\sqrt{c_\parallel/g}$. Putting $P =0$ this structure transforms into the cycloid $\mathbf M = (0,\xi \cos Qy, \eta \sin Qy)$ when $a_z + bM_2^2/4-c_\parallel^2/(2g) = 2 a=0$.
Without clamping $P$ this changes as follows. First of all, we have to consider the possibility of having $M_{z}=M_{z}(y,z)\not=0$ and $P=P(y,z)\not=0$ simultaneously. Then, in accordance with free energy \eqref{F_PM_inh}, these quantities must satisfy the constituent equations 
\begin{subequations}\begin{align}
\big[A - C(\partial _y^2 + \partial _z^2)\big]P
+ 
f_0 \big[M_z(\partial_y M_y) - M_y(\partial_y M_z)\big] &= -\partial _z V, \\
(a_z + bM_y^2 + c_\parallel \partial_y^2 +{g \over 2} \partial_y^4 - 2c_\perp \partial_z^2)M_z 
+ f_0 \big[2 P (\partial_y M_y) + M_y(\partial_y P)\big] &= 0,
\label{}\end{align}\label{equilibrium}\end{subequations}
where $V$ is the electrostatic potential in the system (due to space variations of $P$). Putting $M_z = \eta(y,z) \sin Q y + \theta(y,z)\cos Q y$ and linearizing the above equations we get
\begin{subequations}\begin{align}
\big[A - C (\partial _y^2 + \partial _z^2)\big]P
- f \Big(\eta + {1\over 2Q}(\partial_y \theta)\Big)&= -\partial _z V,\label{P0}\\
(a - c_\parallel \partial_y^2 - c_\perp \partial_z^2 ) \eta -fP&=0,\label{eta0}\\
2(a - c_\parallel \partial_y^2 - c_\perp \partial_z^2 ) \theta 
+ f Q^{-1} ( \partial _y P) &=0.
\label{n}
\end{align}\label{linearized0}\end{subequations}
where $f = f_0 \xi Q $ and $a = [a_{y} + bM_2^2/4-c_\parallel^2/(2g)]/2 $. Since $\eta$, $\theta$ and $P$ are expected to be smoother functions than $\sin Qy$ and $\cos Qy$, higher harmonics have been neglected.

We note that the function $\theta(y,z) $ describes local changes in the phase of the magnetic cycloid which, in accordance that \eqref{n}, are associated with the non-uniformity of the distribution of polarization. In our problem, this distribution is expected to vary at distances much larger than the period of the cycloid $2\pi / Q$. Therefore, the phase of the cycloid remains practically unaltered and actually can be neglected for our purposes ($\theta/\eta \sim k_y /Q \ll 1$, where $k_y$ is the wavevector for the space variations of $P$). Eq. \eqref{n} then can be omitted, and Eqs. \eqref{P0} and \eqref{eta0} reduce to Eqs. (1a) and (1b).


\begin{references}
\bibitem{fnote1}
Real electrodes have finite screening lengths and/or there can be insulating ``dead layers'' in between the ferroelectric-metal interfaces which, mathematically, are equivalent situations \cite{Bratkovsky09}.
\end{references}
\end{document}